\documentclass[12pt,a4paper,leqno]{article}
\usepackage[english]{babel}
\usepackage[latin1]{inputenc}
\usepackage{graphicx}
\usepackage{epsfig}
\usepackage{amsmath, amssymb}
\usepackage{color}
\usepackage{stmaryrd}
\usepackage{fancyhdr}
\usepackage{vmargin}
\setmarginsrb{1.8cm}{1.5cm}{1.8cm}{1.5cm}{0.5cm}{0.5cm}{0.5cm}{0.5cm}

\numberwithin{equation}{section}

\definecolor{rosso}{rgb}{1,0,0}
\definecolor{nero}{cmyk}{0,0,0,1}
\definecolor{blu}{rgb}{0,0,1}

\title{Perturbative Approach on Financial Markets}

\author{\linespread{1} Simone Scotti\footnote{My PhD supervisor
Nicolas Bouleau deserves my sincere gratitude for the constant backing, helpfulness and for the
very relevant contributions brought in this work. I am
grateful to Christophe Chorro, Jean-Francois Delmas, Marzia De Donno, Monique
Jeanblanc, Vathana Ly Vath
and Maurizio Pratelli for their helpful comments, email:
simone.scotti@unito.it; address: Via Real Collegio 30, 10024
Moncalieri (TO) Italy .}
\\ \\ {\it \small  University of Torino and Ecole des
   Ponts - CERMICS}}

\date{ }

\newtheorem{teorema}{Theorem}[section]
\newtheorem{lemma}{Lemma}[section]
\newtheorem{proposizione}{Proposition}[section]
\newtheorem{remarque}{Remark}[section]
\newtheorem{definition}{Definition}[section]

\newtheorem{exemple}{Example}[section]

\begin{document}

\color{nero}

\maketitle

\begin{abstract}
  We study the point of transition between complete and incomplete
  financial models thanks to Dirichlet Forms methods.
  We apply recent techniques, developped by Bouleau, to hedging procedures in order to perturbate parameters and
  stochastic processes, in the case of a volatility parameter fixed
  but uncertain for traders; we call this model Perturbed Black Scholes (PBS) Model.
  We show that this model can reproduce at the same time a smile effect and a bid-ask spread; we
  exhibit the volatility function associated to the local-volatility  model equivalent
  to PBS model when vanilla options are concerned.
  Lastly, we present a connection between Error Theory using Dirichlet Forms and Utility
  Function Theory.


\end{abstract}

\section{Introduction}

In this article, we study the impact of uncertainty of volatility on
the price of vanilla options.

A classical approach consists in perturbating volatility by a
small random variable and in considering the associated truncated
expansion, but this approach encounters difficulties in dealing with
infinite dimensions. An alternative way, based on Dirichlet Forms
(for reference see Albeverio \cite{bib:Albeverio}, Bouleau et al.
\cite{bib:Bouleau-Hirsch} and Fukushima et al.
\cite{bib:Fukushima}), has been suggested by Bouleau \cite{bib:Bouleau-erreur};
this idea yields the same representation
of small perturbations as that of the classical perturbation
approach.

In classical theory of financial mathematics, we assume that all
market securities have a definite price. Indeed, the hypothesis of
completeness of the market (see Lamberton et al. \cite{bib:Lamberton_Lapeyre}) forces
a single price for a contingent claim. If we take  into account an
uncertainty on a parameter, we find that the price of the contingent
claim is not unique but we have many possible prices, therefore we
can reproduce the bid-ask spread by means of a utility function
related to the uncertainty on prices.

If the uncertainty on parameter is small, we may neglect  orders
higher than the second, so we chose to work with Gaussian
distributions. In Error Theory using Dirichlet Forms, we associate
the variance to the ``carré du champ'' operator $\Gamma$ and the
shift to the generator of semigroup $\mathcal{A}$.



Historical Black Scholes model for asset pricing assumes that
the diffusion process for asset price is log-normal with a
constant volatility; however, many works on empirical market data present a skewed
structure of market implied volatilities  with respect to the
strike; this effect is called smile of volatility or volatility
skew (see Renault et al. \cite{bib:Renault-Touzi} and Rubinstein \cite{bib:Rubinstein}).

Implied volatility is convex as a function of the strike and
generally exhibits a slope with respect to the strike at forward
money (see Perignon et al. \cite{bib:Perignon-Villa}); to take this into account, we
propose a new model based on BS model, characterized by an uncertain
volatility parameter, called Perturbed Black-Scholes model. This is
a stochastic volatility model with closed forms for option pricing;
we study some constraints to force a smile on implied volatility and define the local volatility model, by means of its volatility
function, equivalent to the PBS model for vanilla options.

Summarizing, we propose a new financial model for securities
pricing based on the Black Scholes model with a random variable as
volatility, we use a perturbative approach to preserve closed forms
for options prices and greeks; this model permits to reproduce a
smile on implied volatility and generate automatically a bid-ask
spread.

The paper is organized as follows:

In section 2 and 3, we present two ways to perturb the volatility,
the classical approach and the one using Dirichlet Forms. In section
4, we present the PBS model and study the effect of uncertainty
on volatility for an underlying following Black Scholes model
without drift. In section 5, we investigate the relations with the
literature, while
 in section 6, we present an interpretation of the
relative index defined in section 4 by means of utility functions
theory. Finally section 7 resumes and concludes.

\section{Classical Approach}

We consider a function F, which denotes the payoff or the hedging
value of an option and depends on a parameter $\sigma$ (the
volatility for instance), then we consider a perturbation of the
parameter by means of a "normal" distribution and analyze the
impact of this perturbation on F; we suppose that $F \in C^2$ with
respect to the parameter $\sigma$.

This approach may be performed with elementary limit calculations
in a finite dimensional framework:

\subsubsection*{One-dimensional Case}

Assume that $\sigma \in \mathbb{R}$. We model the perturbation on
the parameter $\sigma$ by means of the following transformation:

\begin{displaymath}
\sigma_0 \rightarrow \sigma_0 + \sqrt{\epsilon \gamma} g +
\epsilon a
\end{displaymath}
where $g$ is a random variable, following the centered reduced
normal law and independent of  all the other random variables
present in this financial problem, like the Brownian Motion; $a$ is
a constant and $\epsilon$ is a very small parameter. In other words
we replace the parameter with a random variable with mean $\sigma_0
+ \epsilon a$ and variance $\epsilon \gamma$. We can interpret the
term $\gamma$ as the normalized variance of the variable $\sigma$,
subject to the perturbation; $a$ is the bias of the parameter
induced by the perturbation.

We evaluate the bias and the variance of the pertubation of F,
i.e. $F(\sigma_0+ \sqrt{\epsilon \gamma } g + \epsilon a) -
F(\sigma_0)$.

\begin{eqnarray*}
  \mathbb{E}\left[F(\sigma_0+ \sqrt{\epsilon \gamma } g + \epsilon a) - F(\sigma_0) \right] & = & \mathbb{E}\left[F'(\sigma_0) \left(\sqrt{\epsilon \gamma } g + \epsilon a \right) + \frac{1}{2} F''(\sigma_0) \left(\sqrt{\epsilon \gamma } g + \epsilon a \right)^2 + o(\epsilon) \right]  \\
& = & \epsilon\left\{ a \, \mathbb{E}\left[F'(\sigma_0)\right] +\frac{1}{2} \gamma \, \mathbb{E}\left[F''(\sigma_0)\right] \right\} + o(\epsilon) \\
\mathbb{E}\left[\left(F(\sigma_0+ \sqrt{\epsilon \gamma } g + \epsilon a ) - F(\sigma_0) \right)^2 \right] & = & \mathbb{E}\left[\left(F'(\sigma_0)\right)^2 \left(\sqrt{\epsilon \gamma } g + \epsilon a \right)^2 \right] + o(\epsilon)  \\
& = & \epsilon \gamma
\mathbb{E}\left[\left(F'(\sigma_0)\right)^2\right] + o(\epsilon)
\end{eqnarray*}

Finally we can remark:

\begin{remarque}
The bias and variance have the two following chain rules:

\begin{enumerate}
\item{the bias of the function F, induced by the parameter uncertainty, depends both on the bias and the variance of the
parameter. Indeed, this bias presents two terms. The first
term, given by the bias of the perturbed parameter, is proportional
to the  first derivative. The second term is related
to the convexity of the function F and proportional to the variance
of the parameter; it has a purely probabilistic origin (see
Bouleau \cite{bib:Bouleau-erreur} and \cite{bib:Bouleau-erreur2}).}

\item{the variance of the function F is proportional to the variance of the perturbed
parameter and to the first derivative of the function.}
\end{enumerate}

\end{remarque}

\begin{figure}[h]
  \begin{center}
    \epsfxsize=15cm
    $$
    \epsfbox{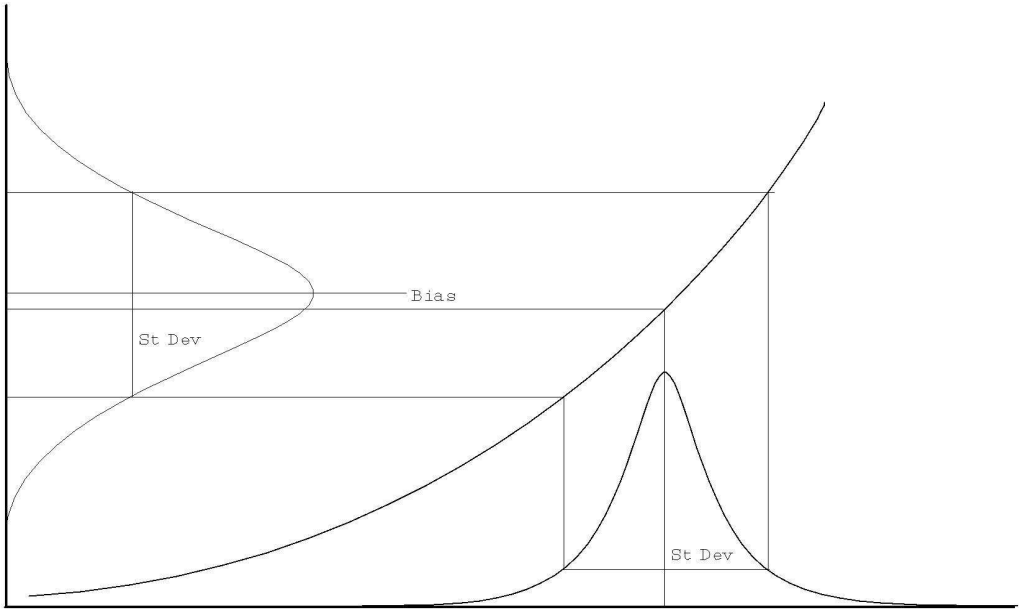}
    $$
        \caption{Impact of uncertainty on a parameter through a non-linear function.}
         \label{impact-convexite}
  \end{center}
\end{figure}

The figure \ref{impact-convexite} resumes the two impacts.

\subsubsection*{Multi-dimensional Case}

We can extend the previous results to the case of a {\it d}-dimensional
parameter; assume $\sigma_0 \in \mathbb{R}^d$ and  $F: \mathbb{R}^d
\rightarrow \mathbb{R}$. In this case the part of the parameter is
expressed as

\begin{displaymath}
\sigma_0 \rightarrow \sigma_0 + \sqrt{\epsilon \gamma} G +
\epsilon A
\end{displaymath}

where

\begin{eqnarray*}
& &  G \sim \mathcal{N}(0, \, \mathbb{I}_d) \\
& &  \gamma \in M(d) \; \;  symmetric \;\, and \;\,
positive \;\, definite \\
& & \gamma^{\frac{1}{2}} \; \; is \; \, a \;\, square \;\,
root \;\, of \;\, \gamma \\
& &  a \in \mathbb{R}^d
\end{eqnarray*}

We suppose again   that $\epsilon \in \mathbb{R}$ is a small
parameter, so that we can approximate the bias and the variance

\begin{eqnarray*}
  \mathbb{E}\left[F(\sigma_0+ \sqrt{\epsilon} \gamma^{\frac{1}{2}} G + \epsilon A ) - F(\sigma_0) \right] & = & \mathbb{E}\left[\displaystyle\sum_{i=1}^d  \frac{\partial F}{\partial x_i}(\sigma_0) \left(\sqrt{\epsilon} \displaystyle\sum_{j=1}^d \gamma^{\frac{1}{2}}_{i, \, j} G_j + \epsilon A_i\right)  \right. \\
 & & + \frac{1}{2} \displaystyle\sum_{i, \, k =1}^d \left. \frac{\partial^2F}{\partial x_i \partial x_k }\right|_{\sigma_0} \left(\sqrt{\epsilon}\displaystyle\sum_{j=1}^d \gamma^{\frac{1}{2}}_{i, \, j} G_j + \epsilon A_i\right)  \\
 & & \left. \times \left(\sqrt{\epsilon}\displaystyle\sum_{l=1}^d \gamma^{\frac{1}{2}}_{k, \, l} G_l + \epsilon A_k\right) + o(\epsilon) \right]  \\
& = & \epsilon \left\{ \vec{A} \; . \; \mathbb{E}\left[\vec{\nabla} F(\sigma_0)\right]  \right. \\
 & & \left. +\frac{1}{2} \displaystyle\sum_{i, \, k=1}^d \gamma_{i, \, k} \, \mathbb{E}\left[\left. \frac{\partial^2F}{\partial x_i \partial x_k }\right|_{\sigma_0}\right] \right\} + o(\epsilon) \\
\mathbb{E}\left[\left(F(\sigma_0+ \sqrt{\epsilon} \gamma^{\frac{1}{2}} G + \epsilon A) - F(\sigma_0) \right)^2 \right] & = & \mathbb{E}\left[\left\{\displaystyle\sum_{i=1}^d \left. \frac{\partial F}{\partial x_i} \right|_{\sigma_0} \left(\sqrt{\epsilon}\displaystyle\sum_{j=1}^d \gamma^{\frac{1}{2}}_{i, \, j} G_j + \epsilon A_i\right) \right\}^2 \right] + o(\epsilon)  \\
& = & \epsilon \; \mathbb{E}\left[ \; ^t\left(\vec{\nabla} F
(\sigma_0)\right) \; \gamma \; \vec{\nabla} F(\sigma_0)\right] +
o(\epsilon)
\end{eqnarray*}
where $\gamma^{\frac{1}{2}}_{k, \, l}$, indicate the element ({\it k l})
of matrix $\gamma^{\frac{1}{2}}$, $A_i$ indicate the $i^{th}$
component of vector $A$ and likewise $G_i$ is the $i^{th}$ row of
random matrix $G$.

Finally we find a similar result  in one-dimensional case:

\begin{remarque}
The bias and variance have the two following chain rules:

\begin{enumerate}
\item{the bias of the function F induced by the uncertainty of the
parameter depends both on the bias and the variance of the
parameter, as a matter of fact this bias presents two terms: a
first term given by the scalar product of the perturbed parameter
bias and the  gradient of function F; the second term is related
to the convexity of the function F and proportional to the
covariance of the parameter $\sigma$;}

\item{the variance of the function F induced by the uncertainty of
the parameter is proportional to the norm of gradient of F
weighted by the  covariance matrix  of the perturbed parameter
$\sigma$.}
\end{enumerate}

\end{remarque}

Several ways are conceivable in order to extend this approach to
infinite dimentional parameters. This is compulsory  since
we are generally concerned with perturbation of a stochastic process.
One way is based on Dirichlet Forms which possess several convenient
featurs for computing the propagation of perturbations.

\section{Dirichlet Forms Approach}

Then we present the Dirichlet Forms approach, we start by
recalling the essential ingredients of this method (see Bouleau
\cite{bib:Bouleau-erreur}).

We define an error structure:

\begin{definition}[Error structure]

An error structure is a term

\begin{displaymath}
  \left( \widetilde{\Omega}, \, \widetilde{\mathcal{F}}, \, \widetilde{\mathbb{P}}, \, \mathbb{D}, \, \Gamma \right)
\end{displaymath}

where

\begin{itemize}

\item{$\left( \widetilde{\Omega}, \, \widetilde{\mathcal{F}}, \,
\widetilde{\mathbb{P}} \right)$ is a probability space;}

 \item{$\mathbb{D}$ is a dense sub-vector space of $L^2\left(
\widetilde{\Omega}, \, \widetilde{\mathcal{F}},
      \, \widetilde{\mathbb{P}}\right)$;}
\item{$\Gamma$ is a positive symmetric bilinear application from
$\mathbb{D} \, \times
    \, \mathbb{D}$ into $L^1 \left( \widetilde{\Omega}, \, \widetilde{\mathcal{F}},
      \, \widetilde{\mathbb{P}} \right)$ satisfying the functional calculus of class
    $\mathcal{C}^1 \cap Lip$, i.e. if F and G are of class $\mathcal{C}^1$ and Lipschitzian, u and
v $\in \mathbb{D}$, we have F(u) and G(v) $\in \mathbb{D}$ and}

\begin{displaymath}
  \Gamma\left[F(u), \, G(v) \right] = F'(u) G'(v) \Gamma[u, \, v] \; \;
  \widetilde{\mathbb{P}} \; a.s.;
  \end{displaymath}

\item{the bilinear form $\mathcal{E}[u, \, v] = \frac{1}{2}
\widetilde{\mathbb{E}}\left[\Gamma[u,
      \, v]\right]$ is closed;}

\item{The constant function 1 belongs to $\mathbb{D}$, i.e. the
error structure is Markovian.}

\end{itemize}

\end{definition}

In mathematical literature the form $\mathcal{E}$ is known as a
"local Dirichlet form" that possesses a ``carré du champ''
operator $\Gamma$.

We recall the definition of sharp operator associated with $\Gamma$

\begin{definition}[Sharp operator]

Let $\left( \widetilde{\Omega}, \, \widetilde{\mathcal{F}}, \,
\widetilde{\mathbb{P}}, \, \mathbb{D}, \, \Gamma \right)$ an error
structure and $\left( \widehat{\Omega}, \, \widehat{\mathcal{F}},
\, \widehat{\mathbb{P}} \right) $ a copy of the probability space
$\left( \widetilde{\Omega}, \, \widetilde{\mathcal{F}}, \,
\widetilde{\mathbb{P}}\right) $. Under the Mokobodzki hypothesis
that the space $\mathbb{D}$ is separable, there exists an operator
sharp $(\,)^{\#}$ with these three properties:

\begin{itemize}

\item{$\forall \, u \in \mathbb{D}$, $u^{\#} \in
L^2(\widetilde{\mathbb{P}} \times \widehat{\mathbb{P}})$;}

\item{$\forall \, u \in \mathbb{D}$, $\Gamma[u] =
\widehat{\mathbb{E}}\left[\left(u^{\#}\right)^2\right]$;}

\item{$\forall \, u \in \mathbb{D}^n$ and $F \in \mathcal{C}^1 \cap
Lip$,  $\left(F(u_1, \, ... \,, \, u_n)\right)^{\#} =
 \sum_{i=1}^n \left(\frac{\partial F}{\partial x_i}\circ u \right) u_i^{\#}  $.}

\end{itemize}

\end{definition}

From a numerical point of view,
the sharp operator is an useful tool to compute $\Gamma$ because the
sharp is linear whereas the carré du champ is bilinear. We can also
associate to the error structure $\left( \widetilde{\Omega}, \,
\widetilde{\mathcal{F}}, \, \widetilde{\mathbb{P}}, \, \mathbb{D},
\, \Gamma \right)$
 a unique
strongly continuous contraction semi-group $(P_t)_{t\geq
  0}$ via Hille Yosida theorem (see Albeverio \cite{bib:Albeverio} pages: 9-11 and 20-26).
This semigroup has a generator $(A, \, \mathcal{D} A)$; it is a
self-adjoint operator that satisfies, for $F \in \mathcal{C}^2$, $u
\in \mathcal{D} A$ and $\Gamma[u] \in L^2(\widetilde{\mathbb{P}})$:

\begin{displaymath}
  A\left[F(u) \right] = F'(u) A[u] + \frac{1}{2} F''(u) \Gamma[u] \; \;
  \widetilde{\mathbb{P}} \; a.s.
\end{displaymath}

The functional calculus extends the ideas of classical Gauss error
theory, the idea is to consider the perturbation as an error.
 In analogy with the classical approach of error theory we associate
the carré du champ operator $\Gamma$ to the
normalized\footnote{the variance divided by the
  small parameter $\epsilon$.} variance of the error, the sharp operator becomes a linear version of the standard deviation of the error.
Similarly, the generator describes the error biases after
normalization (for more details we refers to the book of Bouleau
\cite{bib:Bouleau-erreur} chapters III and V or to
\cite{bib:Bouleau-erreur2}).

\subsection{Perturbation on a parameter}

Now we go back to the study of the impact of a perturbation. To do
so we implement an error structure on the parameter. Therefore we
consider the same function $F \in \mathcal{C}^2$ depending on a
parameter, the volatility for instance. We model the perturbation on
the  parameter with the following transformation:

\begin{displaymath}
  \sigma_0 \rightarrow \sigma = \sigma_0 + X.
\end{displaymath}

Here $X$ is a random variable that represents the error on the
parameter $\sigma$. We consider an error structure associated to $X$ with some hypotheses to
allow the computations.

Assumptions:

\begin{enumerate}
\item{$X \in \mathbb{D}$, $\Gamma[X]$ and $A[X]$ are known;}

 \item{functions $x \mapsto \Gamma[X](x)$ and $x \mapsto A[X](x)$ belong to $L^1 \cap L^2$ are
continuous at
    zero and not vanishing;}
\item{the error structure possesses a gradient operator, allowing
to define the sharp operator which is a special case of gradient
(see Bouleau et al. \cite{bib:Bouleau-Hirsch} chapter V $\S$ 5.2
or Bouleau \cite{bib:Bouleau-erreur} chapter V $\S$ 2 ).}
\end{enumerate}

Denoting $X^{\#}$ the sharp of the parameter $\sigma$, we
have

\begin{eqnarray*}
  F(\sigma)^{\#} & = & F'(\sigma) X^{\#} \\
  \Gamma[F(\sigma)] & = & \left[F'(\sigma)\right]^2 \Gamma[X] \\
  A\left[F(\sigma)\right] & = & F'(\sigma) A[X] + \frac{1}{2} F''(\sigma) \Gamma[X].
\end{eqnarray*}

Following the idea of truncated development, we are interested in
evaluating the bias and the variance at the value $\sigma =
\sigma_0$; i.e. at $X=0$:

\begin{eqnarray*}
  \Gamma[F(\sigma)](\sigma_0) & = & \left[F'(\sigma_0)\right]^2 \Gamma[X]|_{X=0} \\
  A\left[F(\sigma)\right](\sigma_0) & = & F'(\sigma_0) A[X]|_{X=0} + \frac{1}{2} F''(\sigma_0) \Gamma[X]|_{X=0}.
\end{eqnarray*}

 We can remark that we
have found the same chain rules for bias and variance as in the
classical approach; however it is worth notice that we had not to
specify the dimension of the space where $X$ is defined. This dimension
can be finite or infinite, as in thecase of the Wiener space, where we are able to
define  coherent error structures (See Bouleau et al.  \cite{bib:Bouleau-Hirsch} and
\cite{bib:Bouleau-erreur2}). Hence this framework is an extension of the classical approach.

\medskip

Now we present a classical case where we can perform explicit
computation.

\begin{exemple}[Diffusion]
\end{exemple}

\begin{proposizione}
Let $X_t$ be a diffusion process of the form

\begin{equation*}
  dX_t = a(X_t) dB_t + b(X_t) dt
\end{equation*}
where $a(.)$ and $b(.)$ satisfy classical conditions and $B_t$
is a brownian motion. We use this SDE to perturb a parameter
$X_0$, the time t play the role of the small parameter $\epsilon$
 and we are interested in studying the impact of this perturbation on a smooth
 function $F$ depending on the
parameter $X_0$, we have the following bias and variance:

\begin{eqnarray*}
\left. A[F(X)] \right|_{X=X_0} & = & F'(X_0) b(X_0) + \frac{1}{2} F^{\prime \prime}(X_0) a^2(X_0) \\
\left. \Gamma[F(X)] \right|_{X=X_0} & = & \left[F'(X_0)\right]^2
a^2(X_0).
\end{eqnarray*}

The bias is given by the generator of the diffusion and the
variance by the martingale term.

\end{proposizione}

We use a short lemma:

\begin{lemma}
Let $M_t$ a stochastic integral:
\begin{displaymath}
M_t = \int_0^t K_s L_s dB_s
\end{displaymath}
where $K_t$ and $L_t$ are continuous squared integrable processes  and
$N_t$ the following stochastic scaled brownian motion:
\begin{displaymath}
N_t = K_0 L_0 B_t.
\end{displaymath}
Then $N_t$ is an approximation of $M_t$ in $L^2$ when t goes to zero.

\end{lemma}

Proof of lemma: Indeed

\begin{eqnarray*}
  \frac{1}{t} \mathbb{E} \left[\left(M_t - N_t\right)^2\right] & =
  &\frac{1}{t} \int_0^t \left(K_s L_s - K_0 L_0
  \right)^2 ds \\
& \leq & \frac{2}{t} \int_0^t \left\{ \left[K_s\right]^2\left[L_s
-
 L_0
  \right]^2 + \left[K_s -
K_0\right]^2 \left[ L_0
  \right]^2 \right\} ds \rightarrow 0 \; \; \boxempty
\end{eqnarray*}

Proof of proposition 3.1: By It\^{o} formula

\begin{eqnarray*}
  F(X_t) & = & F(X_0) + \int_0^t F'(X_s) a(X_s) dB_s + \int_0^t \left[
  F'(X_s) b(X_s) + \frac{1}{2} F''(X_s) a^2(X_s) \right] ds  \\
& = & F(X_0) + \int_0^t A\left[F(X_s)\right] ds + M_t
\end{eqnarray*}
where $M_t$ is a martingale.

Thanks to the previous lemma, we approximate $M_t$ with $N_t$, given by

\begin{equation*}
  N_t = F'(X_0) a(X_0) B_t
\end{equation*}

Finally we can write the approximate law of $F(X)$ generated by the
perturbation on the parameter $X_0$

\begin{displaymath}
  F(X_t) \approx F(X_0) + t A[F(X_0)] + \sqrt{t} F'(X_0) a(X_0)
  B_1 \; \; \;
\end{displaymath}

where $F(X_0)$, $A[F(X_0)]$ and $F'(X_0)
  a(X_0)$ are $\mathcal{F}_0$-measurable. \hspace{0.5cm} $\boxempty$

\begin{remarque}
We note that the bias and the variance are proportional to
$\epsilon$ and we emphasize that the bias is exactly the generator
of diffusion computed for the function $F$ at the starting point $X_0$,
the variance is the quadratic variation of diffusion associate to
function $F$ at the starting point $X_0$; this example show the relation
between functional analysis and perturbative theory.
\end{remarque}

Finally we can state:

\begin{teorema}[impact of uncertainty]

The impact of uncertainty on the parameter transforms a constant into
a gaussian distribution of the form

\begin{equation}
F(\sigma_0) \rightarrow F(\sigma_0) + \epsilon \;
A[F(\sigma)]|_{\sigma= \sigma_0} + \sqrt{\epsilon} \;
\sqrt{\Gamma[F(\sigma)]|_{\sigma= \sigma_0}} \; G
\end{equation}
where G is an exogenous independent standard Gaussian variable.

\end{teorema}

\begin{remarque}
The bias $\epsilon A[F(\sigma)]|_{\sigma= \sigma_0}$ exists even
though the parameter is unbiased; it suffices that the function F is
non linear.
\end{remarque}

This expansion explains the role of the generator and the carré du
champ operator (see Bouleau \cite{bib:Bouleau-MC}); the theoretical image is
perturbed due to the uncertainty on the parameter. This effect is
small, however  it produces not only a noise but also it alters the
mean.

In this article we study in particular this shift of mean and we
search to reproduce the smile effect by means of this shift.

\section{Perturbed Black Scholes Model}

We start with the classical Black Scholes model (see Black et al.
\cite{bib:Black-Scholes} and Lamberton et al. \cite{bib:Lamberton_Lapeyre}); let
$\left(\Omega, \, \mathcal{F}, \, \mathbb{P}\right)$ the historical
probability space and $B_t$ the associated brownian motion, we
suppose that the dynamic of the risky asset under historical
probability $\mathbb{P}$ is given by the following BS diffusion without drift\footnote{We can remark further in this article that the
presence of a drift term has an impact otherwise from the classical
BS model.}:

\begin{eqnarray*}
dS_t & = & S_t \sigma_0 dB_t \\
S_t & = & S_0 e^{\sigma_0 B_t -\frac{1}{2} \sigma^2_0 t}
\end{eqnarray*}

In this framework, the price of a European vanilla option is well known (see
Lamberton et al.
\cite{bib:Lamberton_Lapeyre}).

The B\&S model present many advantages, in particular the pricing
only depends on volatility and we find closed forms for premium
and greeks of vanilla options; unluckily the B\&S model cannot
reproduce the market price of call options for all strikes at the
same volatility, this effect is called smile.

We propose to consider a perturbation of this model by means of an
error structure on volatility.

We make three hypotheses:

\begin{enumerate}

\item{ the real market follows a B\&S model with fixed and non
perturbed volatility $\sigma_0$ ;}

\item{ the trader has to estimate the volatility, so its
volatility contains intrinsic inaccurancies, we model this ambiguity by means of an
error structure; nonetheless we assume that the stock price $S_t$ is not erroneus. We evaluate the impact of the perturbation,
generated by the
trader mishandling, on the
profit and loss process used by trader to hedge the vanilla option};

\item{ the trader knows this perturbation and he wants to modify the
option prices to take into account the bias induced by the
perturbation on volatility}.

\end{enumerate}

\subsection{``Mismatch'' on trading hedging}

We consider a trader that use an "official" BS asset model in order
to hedge vanilla options; he uses the market price to determine the
"fair" values of parameters in his model (in this case only the
level of flat volatility $\sigma_0$) by inversion of pricing
formula.

The trader finds an observed volatility process $\varsigma_t$,
usually known as implied volatility. He hedges his portfolio
according to his volatility, so the price of an  option, that pays a payoff $\Phi$ is $
F(\varsigma_0, \, x,\, 0)$.

We study the profit and loss process associated to the
hedging position.

The profit and loss process at the maturity of a trader that follows
the strategy associated with his volatility $\varsigma_t$ is given
by:

\begin{equation}\label{equation:P-and-L}
P\&L = F(\varsigma_0, \, x, \, 0) + \int_0^T\frac{\partial
F}{\partial x}(\varsigma_t, \, S_t, \, t) dS_t - \Phi(S_T)
\end{equation}

We make two remarks:

\begin{remarque}
The profit and loss process is stochastic, due to two random
sources:

\begin{itemize}
    \item First of all, the stochastic "real" model since the trader cannot use
    the correct hedging portfolio.
    \item Second, the stochastic process $\varsigma_t$, that can depend
    on a random component independent to the brownian motion $B_t$.
\end{itemize}
\end{remarque}

\begin{remarque}
The profit and loss process must be studied on historical probability
$\mathbb{P}$, therefore the presence of a drift on the BS diffusion
modifies the second term of equation \ref{equation:P-and-L}. Without
drift this term is a martingale and this fact simplifies the
computation. The case of BS model with drift will be dealt in a next
article.
\end{remarque}

In order to analyze the law of P\&L process, it is sufficient to
study the expectation on a class of regular test functions $h(P \& L)$ and
the error on them.

We will come back on the role and choice of a particular function
$h$ in the next subsection.

We suppose by simplicity that the trader volatility $\varsigma_t$ is
a time independent random variable:
\begin{equation}\label{simply-vol-model}
    \varsigma_t = \sigma.
\end{equation}

We define an error structure for the volatility $\sigma$,
therefore this volatility admits the following expansion:

\begin{displaymath}
\sigma_0 \rightarrow \sigma_0 + \epsilon A[\sigma](\sigma_0) +
\sqrt{\epsilon \Gamma[\sigma](\sigma_0)} \widetilde{\mathcal{N}}
\end{displaymath}
where $\widetilde{\mathcal{N}}$ is a standard gaussian variable defined in a space
$(\widetilde{\Omega}, \, \widetilde{\mathcal{A}}, \, \widetilde{\mathbb{P}})$ independent to $\Omega$. Furthermore, if the volatility
has been estimated by means of a statistic on  market data, we can
specify the functional $\Gamma$. In fact, thanks to a result of Bouleau and
Chorro \cite{bib:Bouleau-Chorro}, $\Gamma$ is
related with the inverse of the Fisher information matrix.

We want to estimate the variance and bias error of
$\mathbb{E}\left[h(P \& L)\right]$. To perform the calculus, we
assume that $\sigma = \sigma_0$ is the right value of the random
variable in the sense that $\varsigma_t=\sigma_0$ and $P \& L(\sigma_0) =0$.
We have the following relation for the sharp of volatility:

\begin{displaymath}
    \varsigma_t^{\#} = \sigma^{\#}
\end{displaymath}
Then we can state

 \begin{teorema}
We have the following bias and variance:

\begin{eqnarray}
A[\mathbb{E}[h(P \& L)]] & = &  h'(0) \Upsilon_1^{BS}(\sigma_0) +
\frac{1}{2}h''(0)\Upsilon_2^{BS}(\sigma_0) \label{A-P-and-L-BS} \\
    \Gamma \left[\mathbb{E}\left[h\left( P\&L \right)\right]
    \right] & = &  [h'(0)]^2 \; \Lambda^{BS}(\sigma_0) \label{gamma-P-and-L-BS}
\end{eqnarray}
where
\begin{eqnarray*}
\Upsilon_1^{BS}(\sigma_0) & = &
\left\{\frac{\partial F}{\partial \sigma}(\sigma_0, x, 0) A[\sigma] (\sigma_0) +
\frac{1}{2} \frac{\partial^2F}{\partial \sigma^2}(\sigma_0, x, 0) \Gamma[\sigma](\sigma_0) \right\} \\
\Upsilon_2^{BS}(\sigma_0) & = & \left\{\left[\frac{\partial
F}{\partial \sigma}(\sigma_0, x, 0) \right]^2 + \sigma_0^2
\int_0^T \mathbb{E} \left[S_t^2 \left(\frac{\partial^2F}{\partial
\sigma \partial x}(\sigma_0, S_t, t) \right)^2 \right] dt \right\}
\Gamma[\sigma](\sigma_0) \\
\Lambda^{BS}(\sigma_0) & = & \left\{\mathbb{E} \left[\frac{\partial
F}{\partial \sigma}(\sigma_0, x,
    0)\right] \right\}^2 \Gamma[\sigma](\sigma_0)
\end{eqnarray*}
and we have the following truncated expansion:

\begin{equation}\label{equation:rel_h}
  \mathbb{E}\left[h(P\&L)\right] \approx \epsilon \, h'(0) \, \Upsilon_1^{BS}(\sigma_0) + \epsilon
   \,
\frac{1}{2}h''(0) \,\Upsilon_2^{BS}(\sigma_0) + \sqrt{\epsilon \,
\left[h'(0) \right]^2 \,
  \Lambda^{BS} } \; \widetilde{\mathcal{N}}(0, \, 1)
  \end{equation}

\end{teorema}

Proof:

We start with the study of the variance. A computation yields

\begin{eqnarray*}
    \left(\mathbb{E}\left[h\left( P\&L \right)\right]\right)^{\#} & = &
    \mathbb{E}\left[h'\left(P\&L \right) \left(\frac{\partial F}{\partial \sigma}(\sigma_0, x,
    0) + \int_0^T \frac{\partial^2F}{\partial \sigma \partial x}(\sigma_0, S_s, s) dS_s \right) \sigma^{\#}
  \right]
\end{eqnarray*}
thus the quadratic error is equal to:

\begin{eqnarray*}
    \Gamma \left[\mathbb{E}\left[h\left( P\&L \right)\right]
    \right] & = & [h'(0)]^2 \left\{\mathbb{E} \left[\frac{\partial F}{\partial \sigma}(\sigma_0, x,
    0) + \int_0^T \frac{\partial^2F}{\partial \sigma \partial x}(\sigma_0, S_s, s)
    dS_s\right] \right\}^2 \Gamma[\sigma](\sigma_0)
\end{eqnarray*}
and the second term vanishes since the stock price is a martingale, here the
hypothesis on drift is crucial.

The study of the bias is more complicated, we start with the remark that the bias is
a linear operator:








\begin{displaymath}
 A \left[\mathbb{E}\left[h(P \& L)\right]\right] = \mathbb{E}\left[A[h(P \& L)]\right] =
  \mathbb{E} \left[ h'(P \& L) A\left[ P \& L  \right] + \frac{1}{2} h^{\prime
    \prime}(P \& L)\Gamma \left[P \& L\right]\right]
  \end{displaymath}

We study the two terms separately; for the first, we find the expectation of quadratic error in the case
$\varsigma = \sigma_0$:

\begin{eqnarray*}
        \mathbb{E}\left[\Gamma \left[P \& L \right]\right] & = & \left\{ \mathbb{E} \left[
 \left( \frac{\partial F}{\partial \sigma}(\sigma_0, x, 0)\right)^2 \right]  \right. \\
& & \left. \int_0^T \mathbb{E} \left[ \sigma_0^2 S_s^2 \left(
\frac{\partial^2F}{\partial \sigma \partial x}(\sigma_0, S_s, s)
\right)^2 ds \right]
 \right\} \Gamma \left[ \sigma \right] \left( \sigma_0 \right)
\end{eqnarray*}

We study the bias operator; we must evaluate the expectation of
bias of profit and loss process, and we find the following result always in the case $\varsigma = \sigma_0$:

\begin{displaymath}
\mathbb{E}[A[P \& L]] =  \frac{\partial F}{\partial
\sigma}(\sigma_0, x, 0) A[\sigma] (\sigma_0) + \frac{1}{2}
\frac{\partial^2F}{\partial \sigma^2}(\sigma_0, x, 0)
\Gamma[\sigma](\sigma_0)
\end{displaymath}

Finally the bias of expectation of a function of profit and loss
process:

\begin{eqnarray*}
A[\mathbb{E}[h(P \& L)]]    &  = & h'(0) \left\{\frac{\partial F}{\partial \sigma}(\sigma_0, x, 0) A[\sigma] (\sigma_0) + \frac{1}{2} \frac{\partial^2F}{\partial \sigma^2}(\sigma_0, x, 0) \Gamma[\sigma](\sigma_0) \right\}  \\
& & + \frac{1}{2} h''(0) \left\{\left[\frac{\partial F}{\partial \sigma}(\sigma_0, x, 0) \right]^2 + \right. \\
& & \left. +\sigma_0^2 \int_0^T \mathbb{E} \left[S_t^2
\left(\frac{\partial^2F}{\partial \sigma \partial x}(\sigma_0,
S_t, t) \right)^2 \right] dt \right\} \Gamma[\sigma](\sigma_0)
\end{eqnarray*}

The proof ends with the truncated expansion that is a
consequence of the error theory using Dirichlet Forms (see Bouleau
\cite{bib:Bouleau-erreur3} and \cite{bib:Bouleau-erreur4})

In order to interpret this result in finance, we consider that the trader
knows the presence of errors in his procedure and wants to
neutralize this effect.

We associate:

\begin{itemize}
\item the variance of $h(P \& L )$ process  to the bid-ask spread
of options;

\item the bias of $h(P \& L)$ process to a shift of prices of options
asked by the trader to the buyer.
\end{itemize}

\begin{figure}[h!]
  \begin{center}
    \epsfxsize=13cm
    $$
    \epsfbox{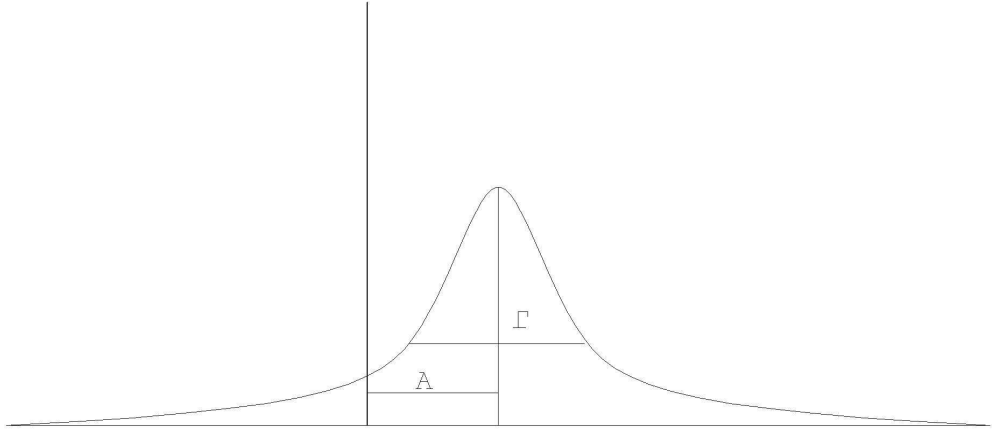}
    $$
    \caption{Impact of ambiguity: the Dirac distribution of price X becomes a continuous
      distribution; the mean shifts of $\epsilon \; A[X]$ and the variance is $ \epsilon \;
      \Gamma[X]$.} \label{fig:error}
  \end{center}
\end{figure}

Indeed in the classical theory of financial mathematics we assume
that all market securities have a single  price, with the
probability theory language we can associate at any derivative
securities a Dirac distribution for its price. If we take into
account uncertainty on volatility, we have found that the price of
the contingent claim is not unique but we have many possible prices;
thus the Dirac distribution changes into a continuous distribution,
characterized by a variance and a shift of the mean with respect to
the previous Dirac distribution (see figure \ref{fig:error}) .

\begin{teorema}
If the perturbation in the volatility is small, we can neglect
orders higher than the second, so we always work with Gaussian
distributions; the trader must modify his prices in order to take
into account the two previous effects, namely the variance and the
bias, then he fixes a supportable risk probability $\alpha < 0.5$
and accepts to buy the option at the price

\begin{displaymath}
(\text{Bid Premium}) = (\text{BS Premium}) + \epsilon \; A
\left[\mathbb{E}[h(P \& L)] \right]  + \sqrt{\epsilon \; \Gamma
\left[\mathbb{E}[h(P \& L)]\right]}\; \mathcal{N}_{\alpha}
\end{displaymath}

where $\mathcal{N}_{\alpha}$ is the $\alpha$-quantile of the reduced
normal law. Likewise, the trader accepts to sell the option at the
price

\begin{displaymath}
(\text{Ask Premium}) = (\text{BS Premium}) + \epsilon \; A
\left[\mathbb{E}[h(P \& L)]\right] + \sqrt{\epsilon \; \Gamma
\left[\mathbb{E}[h(P \& L)]\right]} \; \mathcal{N}_{1-\alpha}
\end{displaymath}

\end{teorema}

\begin{remarque}
We remark that the two previous prices are symmetric, since
$\mathcal{N}_{\alpha} + \mathcal{N}_{1- \alpha}= 0$; therefore  the
mid-premium is

\begin{displaymath}
(\text{Mid Premium}) = (\text{BS Premium}) + \epsilon \; A
\left[\mathbb{E}[h(P \& L)]\right]
\end{displaymath}

We emphasize that with our model we can reproduce a bid-ask spread
and we can associate its width to the trader's risk aversion (the
probability $\alpha$) and the volatility uncertainty (the term
$\sqrt{\epsilon \; \Gamma \left[\mathbb{E}[h(P \& L)]\right]}$).

 In the rest of this article we work directly with
the mid premium, but all results represent the center of a normal
distribution, in order to reproduce the bid and the ask premium we
need to specify the probability $\alpha$.

\end{remarque}

We conclude this subsection with a remark.

\begin{remarque}
The presence of the perturbation induces a problem on the
completeness of the market, the market with perturbed volatility is
not complete, since the volatility depends on a second random source
orthogonal to $\Omega$, and the presence of a bid-ask spread is a
direct consequence of this fact; on the other hand the enforcement
that $\sigma$ to be equal $\sigma_0$ cancels the impact of the
second random source. This apparent contradiction is due to the fact
that an argument acts precisely at the boundary between complete and
incomplete markets.
\end{remarque}

\subsection{Role and choice of h functions and relative index}

In this subsection we discuss on the choice of function $h$ in
equation (\ref{equation:rel_h}), because function h defines the
magnitude of correction on prices; this choice becomes simpler since
we have to specify only the first and second derivative
in zero; therefore we can consider that the function h is a parabola
that passes through the origin. Owing to the two degrees of freedom
associated with $\epsilon$ and $\alpha$, we can take $h'(0) = 1$:
this is easy to understand from the economical point of view because
the trader wants to balance his portfolio, i.e. the $P\&L$ process.
If we look at equation (\ref{equation:rel_h}) we find that the
choice of $h'(0) = 1$ defines completely the term of variance. Since
the second derivative of h has an impact only on the bias and the
coefficient $\Upsilon_2^{BS}(\sigma_0)$ is positive,  this impact
 is a shift of the mean, as in the following figure.

\begin{figure}[h!]
  \begin{center}
    \epsfxsize=12cm
    $$
   \epsfbox{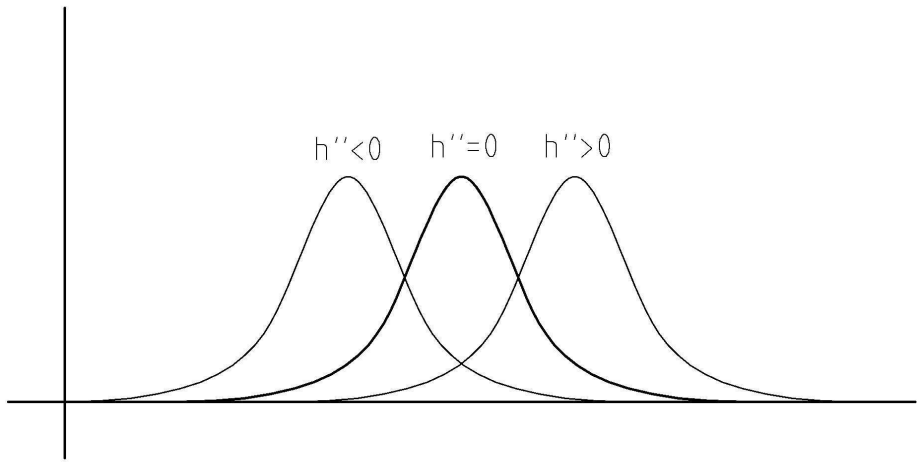}
    $$
    \caption{Impact of ambiguity: the convexity (resp. concavity) of function h
      raises (resp. reduces) the mean of prices but leaves the variance unchanged.}
  \end{center}
\end{figure}

We suggest to interpret this impact as an asymmetry of the balance
between supply and demand. Indeed, if $h''(0)=0$ we find that  the
function h is the identity: this means that the trader uses directly
the process of profit and loss, and the bias is ``neutral'', i.e.
that we find the same result if we consider the buyer's point of
view (it is enough to take minus identity function as h). A surplus
of the demand of an option with respect to the supply induces a
raising of the prices: this is the classical case of market where
banks sell options and
  private investors buy.
We model this perturbation with a positive second derivative for h
and we consider that if $h''(0)>0$ (resp. $h''(0)<0$) the demand
(resp. supply) exceeds the supply (resp. demand).

 We define the
following index of asymmetry of balance between supply and demand:

\begin{displaymath}
  r_{S / D} = \frac{h''(0)}{h'(0)}
\end{displaymath}

We can identify this index by means of the classical utility theory:
if we interpret $h$ as a utility function, $r_{S / D}$ is known as
the absolute index of $h$. In the next part we study the bias of
profit and loss process in the case of a call option; to simplify
matters, we suppose that $h$ is the identity function i.e. $r_{S /
D}=0$, but before we must introduce an other index, very important
in the continuation of this article.

We concentrate our attention on a problem; after the perturbation of
a parameter $\sigma_0$, we have:

\begin{displaymath}
  \sigma_0 \rightarrow \sigma = \sigma_0 + \epsilon \, A[\sigma] +
  \sqrt{\epsilon \, \Gamma[\sigma]} \mathcal{N} \; \; with \; \;
  \mathcal{N} \sim N(0, \, 1)
\end{displaymath}

However $\epsilon$ is generally unknown: the Error Theory via
Dirichlet Forms cannot define this parameter. In order to deal with
this question we propose to renormalize this problem; we consider
the ratio between the bias and the variance, since the variance is
almost surely strictly positive, therefore that the dependence on
$\epsilon$ is cancelled:

\begin{equation*}
\frac{Bias \; X}{Variance \; X} = \frac{ \epsilon A[X]}{\epsilon
\Gamma[X]} = \frac{A[X]}{\Gamma [X]}
\end{equation*}

This ratio is not homogeneous, because the generator is linear
and the operator ``carre du champs'' is bilinear, so we define a
relative index by:

\begin{equation}
  r_r(X) = 2\frac{X A[X]}{\Gamma[X]}
\end{equation}

The factor 2 will be justified in section \ref{app:utility}. where
we show a relation between Dirichlet forms and utility theory, since
we can interpret $r_r(X)$ as a relative index of an exogenous
utility function.

\subsection{Call options case}

We concentrate on call option and we study the bias and its
derivatives in order to determine some sufficient condition to
force the presence of a smile on implied volatility.

 We know the premium of a call option (see Lamberton et al.
\cite{bib:Lamberton_Lapeyre}) with strike $K$ and spot value $x$,
and its hedging strategy:

\begin{eqnarray*}
C(\sigma_0, \, x, \, 0) & = &F(\sigma_0, \, x, \, 0) = x\mathcal{N} (d_1) - K \mathcal{N}(d_2) \\
Delta & = & \frac{\partial F}{\partial x}(\sigma_0, \, x, \, 0) =
\mathcal{N} (d_1)
\\
where & & d_1 = \frac{\ln x - \ln K + \frac{\sigma_0^2}{2} T
}{\sigma_0 \sqrt{T} } \; \; and \; \; d_2 = d_1 - \sigma_0 \sqrt{T}.
\end{eqnarray*}

The following results are classical (see \cite{bib:Lamberton_Lapeyre}):

\begin{eqnarray}
\frac{\partial F}{\partial \sigma_0}(\sigma_0, \, x, \, 0) & = & x\sqrt{T} \frac{e^{-\frac{1}{2} d_1^2}}{\sqrt{2 \pi}} \nonumber \\
 \frac{\partial^2F}{\partial \sigma_0^2}(\sigma_0, \, x, \, 0) & = & \frac{x
   \sqrt{T}}{\sigma_0} \frac{e^{-\frac{1}{2} d_1^2}}{\sqrt{2 \pi}} d_1 d_2 \label{eqn_deriv_K2} \\
\frac{\partial^2F}{\partial K^2}(\sigma_0, \, x, \, 0) & = &
\frac{x}{K^2 \sigma_0 \sqrt{T}} \frac{e^{-\frac{1}{2}
d_1^2}}{\sqrt{2 \pi}}. \nonumber
\end{eqnarray}

Then the bias of the call premium is given by

\begin{equation}
A[C]|_{\sigma= \sigma_0} = x \frac{e^{-\frac{1}{2} d_1^2}}{\sqrt{2
\pi}} \left\{ A\left[\sigma \sqrt{T}\right]|_{\sigma= \sigma_0} +
 \frac{d_1 d_2}{2 \sigma_0 \sqrt{T}} \Gamma\left[\sigma \sqrt{T}\right]|_{\sigma= \sigma_0} \right\}.
\end{equation}

We can compute the first derivative with respect to the strike:

\begin{eqnarray*}
\left. \frac{\partial A[C]}{\partial K}\right|_{\sigma= \sigma_0} &
= & \frac{x}{K \sigma_0 \sqrt{T}} \frac{e^{-\frac{1}{2}
  d_1^2}}{\sqrt{2 \pi}} \left\{ d_1 A\left[\sigma \sqrt{T}\right]|_{\sigma= \sigma_0} - \frac{d_1 + d_2 -
  d_1^2 d_2}{2 \sigma \sqrt{T}} \Gamma\left[\sigma \sqrt{T}\right]|_{\sigma= \sigma_0} \right\} \\
& = & \frac{d_1 A[C]|_{\sigma= \sigma_0}}{K \sigma_0 \sqrt{T}} -
\frac{x}{2 K \sigma_0^2 T} \frac{e^{-\frac{1}{2} d_1^2}}{\sqrt{2
\pi}} (d_1 + d_2) \Gamma\left[\sigma \sqrt{T}\right]|_{\sigma=
\sigma_0}
\end{eqnarray*}

We find that the first derivative vanishes at the forward
money\footnote{In interest-free case the forward money is for K =
x, in this case we have $d_1 = - d_2$.} if and only if the bias of
call vanishes at the same strike.

\begin{eqnarray*}
  \left. A[C] \right|_{K=x, \; \sigma= \sigma_0} & = & x \frac{e^{-\frac{\sigma_0^2 T}{8}}}{\sqrt{2 \pi}}\left\{ A\left[\sigma \sqrt{T}\right]|_{\sigma= \sigma_0} -
    \frac{\sigma_0 \sqrt{T}}{8} \Gamma\left[\sigma \sqrt{T}\right]|_{\sigma= \sigma_0} \right\} \\
   \left. \frac{\partial A[C]}{\partial K} \right|_{K=x, \; \sigma= \sigma_0} & = & \frac{1}{2x} \left. A[C] \right|_{K=x, \; \sigma= \sigma_0}
\end{eqnarray*}

We can remark that the bias and its first derivative are
positive (resp negative) at the money if and only if

\begin{equation}\label{rel_RR_bias}
  \left. r_r^{BS}\left(\sigma \sqrt{T}\right) \right|_{\sigma= \sigma_0} =\left. 2 \sigma_0 \sqrt{T} \frac{A\left[\sigma \sqrt{T}\right]}{\Gamma\left[\sigma
  \sqrt{T}\right]} \right|_{\sigma= \sigma_0} > \, (resp. \; <) \, \frac{\sigma_0^2 T}{4}
  \end{equation}

We find three cases:

\begin{enumerate}
\item{ if $r_r\left(\sqrt{\sigma_0^2 T} \right) < \frac{1}{4}\sigma_0^2 T $
    , then the bias of call and his first derivative  are negative at the money.}
\item{ if $r_r\left(\sqrt{\sigma_0^2 T} \right) = \frac{1}{4}\sigma_0^2 T$
    , then the bias of call and his first derivative  vanish at the money.}
\item{ if $r_r\left(\sqrt{\sigma_0^2 T} \right) > \frac{1}{4}\sigma_0^2 T$
    , then the bias of call price and his first derivative  are positive at the money.}
\end{enumerate}

\begin{remarque}

  This bound increases with maturity; if we suppose a constant relative index (CRI) we can
   define a bound on maturity $\mathcal{T}_{bias}$.

  Therefore if we study an option with maturity smaller (resp. greater) than
  $\mathcal{T}_{bias}$ we have that the bias associated to the hedging "profit and
  loss" process and his first derivative are positive (resp. negative).

  \end{remarque}

We calculate the second derivative:

\begin{displaymath}
\left. \frac{\partial^2 A[C]}{\partial K^2}\right|_{\sigma=
\sigma_0} = \frac{d_2}{K \sigma_0 \sqrt{T}} \left. \frac{\partial
  A[C]}{\partial K} \right|_{\sigma= \sigma_0}- \frac{x}{K^2 \sigma_0^2 T}
\frac{e^{-\frac{1}{2} d_1^2}}{\sqrt{2 \pi}} \left\{ A\left[\sigma
\sqrt{T}\right] |_{\sigma= \sigma_0} + \frac{ d_1^2 + 2d_1 d_2
  -2}{2 \sigma \sqrt{T}} \Gamma\left[\sigma \sqrt{T}\right]|_{\sigma= \sigma_0} \right\}
\end{displaymath}

and we evaluate the second derivative at the forward money

\begin{equation*}
\left. \frac{\partial^2 A[C]}{\partial K^2}\right|_{K=x, \; \sigma=
\sigma_0}
 = - \frac{1}{K} \frac{e^{-\frac{\sigma_0^2 T}{8 }}}{\sqrt{2 \pi}} \left\{ \left( \frac{1}{4} +
    \frac{1}{\sigma_0^2 T} \right)
    A\left[\sigma \sqrt{T}\right]|_{\sigma= \sigma_0} -
    \left[ \frac{\sigma_0^2 T}{32} + \frac{1}{8} + \frac{1}{\sigma_0^2 T}\right]
    \left. \frac{\Gamma\left[\sigma \sqrt{T}\right]}{\sigma \sqrt{T}} \right|_{\sigma= \sigma_0} \right\}.
\end{equation*}

If we force the bias of the call to be convex, we find:

\begin{equation}\label{rel_RR_bias-2-derivative}
\left. r_r^{BS}\left(\sigma \sqrt{T}\right)\right|_{\sigma=
\sigma_0} = \left. 2 \sigma \sqrt{T} \frac{A\left[\sigma
\sqrt{T}\right]}{\Gamma\left[\sigma
  \sqrt{T}\right]}\right|_{\sigma= \sigma_0} <
 \frac{\sigma_0^4 T^2 + 4 \sigma_0^2 T + 32}{4 \sigma_0^2 T + 16} =
 \Theta \left(\sigma_0 \sqrt{T}\right).
\end{equation}

\begin{remarque}

  Previous bound $\Theta(x)$ is strictly positive, decreasing in $[0,
  \, 4\sqrt{2}]$ and increasing if $x> 4 \sqrt{2}$, and we find that $\Theta(4 \sqrt{2}) =
  \frac{7}{2} \sqrt{2} - 3$.

  If the relative index is constant (CRI) we have an always convex bias if $r_r<
  \Theta(4 \sqrt{2})$
  \end{remarque}

In the previous relations we note that the constraint depends on the
volatility by means of cumulated volatility $\sigma_0 \sqrt{T}$. For
more generality in this study we can assume that the erroneous
parameter is not the volatility but the cumulated variance $\int_0^T
\sigma_0^2(s) ds$ that appears in general Black \& Scholes model
when the volatility is deterministic but depends on time.





Now we study the evolution of slope as a function of maturity at the money

\begin{eqnarray*}
 \left. \frac{\partial A }{\partial T} \right|_{\sigma= \sigma_0}  &=& \frac{x}{2 T }
  \frac{e^{-\frac{d_1^2}{2}}}{\sqrt{2 \pi}} \left\{ (1 + d_1 d_2) A\left[\sigma
    \sqrt{T}\right] |_{\sigma= \sigma_0} + \frac{4d_1^2 d_2^2 -3\sigma_0^2 T - (d_1+d_2)^2}{8 \sigma_0 \sqrt{T}}\Gamma\left[\sigma\sqrt{T} \right] |_{\sigma= \sigma_0} \right\} \\
  \left. \frac{\partial A }{\partial T} \right|_{K=x, \; \sigma = \sigma_0} &=&
   \frac{x e^{-\frac{\sigma_0^2 T}{8}}}{8 T \sqrt{2 \pi}}
  \left\{\left(4 - \sigma_0^2 T
    \right) A\left[\sigma \sqrt{T}\right]|_{\sigma= \sigma_0} +\frac{\sigma_0 \sqrt{T}}{8} \left(\sigma_0^2 T - 12\right) \Gamma\left[\sigma\sqrt{T} \right]|_{\sigma= \sigma_0}\right\}
\end{eqnarray*}

\begin{eqnarray*}
  \left. \frac{\partial^2 A }{\partial K \partial T} \right|_{\sigma= \sigma_0}&=& -
  \frac{x}{2 \sigma_0
  T^{\frac{3}{2}} K} \frac{e^{-\frac{d_1^2}{2}}}{\sqrt{2 \pi}} \left\{d_2 (1 - d_1^2)
  A\left[\sigma \sqrt{T}\right]|_{\sigma= \sigma_0} + \right. \\
  & & + \left. d_1
  \frac{4 d_1^3 d_2^2 - 3 d_1 \sigma_0^2 T + (d_1+d_2)(4 - 9d_1d_2 -d_1^2)}{8 \sigma_0 \sqrt{T}}
  \Gamma\left[\sigma\sqrt{T} \right]|_{\sigma= \sigma_0} \right\} \\
  \left. \frac{\partial^2 A }{\partial K \partial T} \right|_{K=x, \; \sigma = \sigma_0} &=&
   - \frac{e^{-\frac{\sigma_0^2 T}{8}}}{16 T \sqrt{2 \pi} }
  \left\{\left(\sigma_0^2 T -4
    \right) A\left[\sigma \sqrt{T}\right]|_{\sigma= \sigma_0} - \sigma_0 \sqrt{T}
    \frac{\sigma_0^2 T - 12}{8} \Gamma\left[\sigma\sqrt{T} \right]|_{\sigma= \sigma_0} \right\}
\end{eqnarray*}

to find a slope that increases with increasing maturity we have to impose:

\begin{equation}\label{slope-time}
    r_r\left(\sigma \sqrt{T} \right)|_{\sigma= \sigma_0} \geq \frac{\sigma_0^2
  T}{4} \; \frac{\sigma_0^2 T - 12 }{ 4- \sigma_0^2 T} \; \; \;
\; \, with \; \, \; \sigma_0^2 T < 4
\end{equation}

We study the evolution of smile as a function of maturity.

\begin{equation*}
  \left. \frac{\partial^3 A }{\partial K^2 \partial T} \right|_{K=x, \; \sigma = \sigma_0}
  = \frac{e^{-\frac{\sigma_0^2 T}{8}}}{x \sigma_0^2 T^2 \sqrt{2 \pi} } \left\{
  \frac{ 16 + \sigma_0^4 T^2}{32} A\left[\sigma
  \sqrt{T}\right]|_{\sigma= \sigma_0} - \frac{ \sigma_0^2 T \left(\sigma_0^2 T -4\right)^2
  +128}{256} \left. \frac{\Gamma\left[\sigma\sqrt{T}
  \right]}{\sigma \sqrt{T}} \right|_{\sigma= \sigma_0} \right\}
\end{equation*}

This term is positive if and only if:

\begin{equation}\label{smile-time}
    r_r\left(\sigma \sqrt{T} \right) |_{\sigma= \sigma_0} > \frac{1}{4} \; \frac{ \sigma_0^2 T
      \left(\sigma_0^2 T -4\right)^2 +128 }{ 16 + \sigma_0^4 T^2 }
\end{equation}







\subsection{Example}

Now we study a particular case, in order to reproduce the smile of
volatility present in market data.

\begin{teorema}

We fix the relative index at

\begin{equation}\label{eqn:particular-case}
  r_r^{BS}(\sigma \sqrt{T})|_{\sigma= \sigma_0} =  \, \frac{\sigma_0^2 T}{4}
  \end{equation}

This choice fix the values of the bias and its derivatives at the
money, in particular this choice force the bias and its first
derivative to be zero at the money (see equation \ref{rel_RR_bias});
the second derivative becomes positive at the money thanks to
equation \ref{rel_RR_bias-2-derivative} for any maturity $T$.
Therefore the bias is strictly convex around the money and vanishes
at the money, therefore it is positive in a neighbourhood of the
money. Now if the bias vanishes at the money the implied ATM
volatility is $\sigma_0$, but, since the bias is positive around,
the implied volatility becomes greater than $\sigma_0$ around the
money\footnote{We recall that the vega is positive for call
options.}; Finally we have reproduced a smile effect around the
money.

\end{teorema}

\begin{figure}[h!]
  \begin{center}
    \epsfxsize=14cm
    $$
   \epsfbox{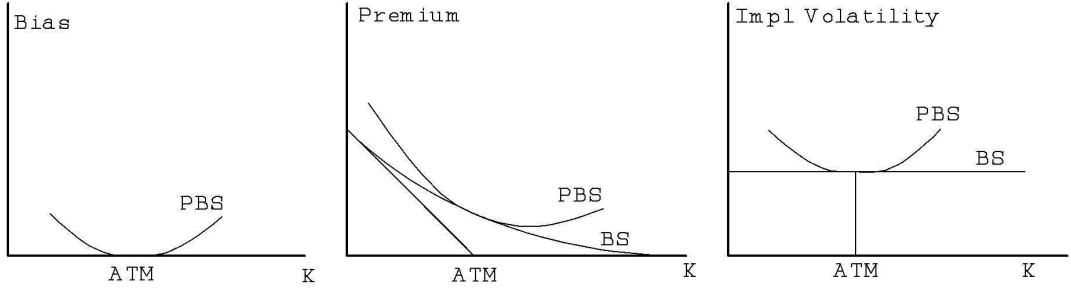}
    $$
    \caption{Representation of the smile in PBS model.}
  \end{center}
\end{figure}

\subsection{Dupire formula and Implicit Local Volatility Model}

In this section we want to specify the Local Volatility Model
equivalent to Perturbed Black Scholes Model. We know that the
knowledge of prices of options for all strikes and maturities
defines a single local volatility model that reproduces these
prices; the Dupire formula (see Dupire \cite{bib:Dupire}) defines the local
volatility function:

\begin{equation}
  \sigma^2_{imp}(T, \, K) = \frac{ \frac{\partial C}{\partial T}}{\frac{1}{2}
  K^2 \frac{\partial^2 C }{\partial K^2}}
\end{equation}

Our model is a perturbation of  BS model, so we can consider the
following expansion:

$$
\begin{array}{rccl}
  & & BS \, Term & Perturbation \\
  & & & \\
  C(\sigma, \, x, \, K, \, t, \, T) & = & C(\sigma_0, \, x, \, K, \, t, \, T ) +
  & \epsilon \,
  A[C](\sigma, \, x, \, K, \, t, \, T)|_{\sigma= \sigma_0} \\
  & & & \\
  \displaystyle \frac{\partial C}{ \partial T}(\sigma, \, x, \, K, \, t, \, T) & = & \displaystyle
  \frac{\partial C}{ \partial T}(\sigma_0, \, x, \, K, \, t, \, T) + & \epsilon \, \displaystyle
  \left. \frac{\partial A[C]}{ \partial T}(\sigma, \, x, \, K, \, t, \, T)\right|_{\sigma= \sigma_0} \\
  & & & \\
\displaystyle \frac{\partial^2 C}{ \partial K^2}(\sigma, \, x, \, K,
\, t, \, T) & = & \displaystyle
  \frac{\partial^2 C}{ \partial K^2}(\sigma_0, \, x, \, K, \, t, \, T) + & \epsilon \, \displaystyle
  \left. \frac{\partial^2 A[C]}{ \partial K^2}(\sigma, \, x, \, K, \, t, \, T)\right|_{\sigma= \sigma_0}
\end{array}
$$

But in fact, if $\epsilon$ vanishes, the model is a Black Scholes model with
volatility $\sigma_0$.
Thanks to equation (\ref{eqn_deriv_K2}), we can rewrite:

\begin{equation*}
  \sigma^2_{imp}(T, \, K) \approx \sigma_0^2 + \frac{\epsilon }{\frac{1}{2} K^2
  \frac{\partial^2 C }{\partial K^2}}\left[ \left. \frac{\partial
        A[C]}{\partial T}\right|_{\sigma= \sigma_0} - \frac{1}{2} K^2 \sigma_0^2
        \left. \frac{\partial^2 A[C]}{ \partial K^2}\right|_{\sigma= \sigma_0} \right]
\end{equation*}

\begin{teorema}

The local volatility model equivalent to the PBS model for vanilla
options,  has the following local volatility function:

\begin{equation}
  \sigma^2_{imp}(T, \, K) \approx \sigma_0^2 \left\{ 1 + \frac{\epsilon}{2}
  \left[ 4 \frac{A\left[\sigma \sqrt{T}\right] |_{\sigma= \sigma_0}}{\sigma_0 \sqrt{T}} -
  \left[\sigma_0^2 T +
       2 - \frac{4 \ln\left(\frac{x}{K}\right)^2}{\sigma_0^2 T} \right] \frac{\Gamma\left[\sigma
         \sqrt{T}\right] |_{\sigma= \sigma_0} }{\sigma_0^2 T} \right] \right\}
\end{equation}

\end{teorema}

\begin{remarque}
  Local volatility $\sigma(T, \, K)$ has a minimum at forward money $K=x$, and
  present a logarithmic behavior as K approaches zero and infinity.

  We must preserve the positivity of the square of volatility, so we fix the following
  constraint:

  \begin{equation*}
       \sigma_0^2 T +2 - 2 r_r\left(\sigma \sqrt{T} \right)|_{\sigma= \sigma_0}
      <  \left\{\epsilon \frac{\Gamma\left[\sigma
         \sqrt{T}\right]|_{\sigma= \sigma_0}}{\sigma_0^2 T} \right\}^{-1}
    \end{equation*}
\end{remarque}

\begin{figure}[h!!]
  \begin{center}
    \epsfxsize=10cm
    $$
   \epsfbox{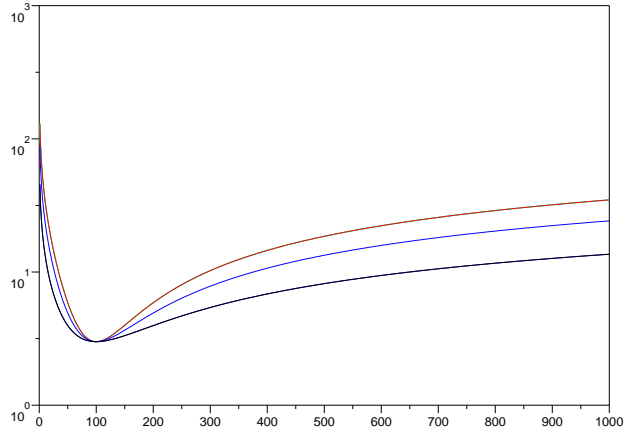}
    $$
    \caption{Examples of local BS and PBS variances (square of volatilities)  for ATM = 100, $\sigma \sqrt{T} = 30 \%$,
      $\epsilon = 0.01$,
      $\Gamma[\sigma \sqrt{T}] = \sigma^2 T$ and $r_r=0$.}
  \end{center}
\end{figure}

\section{Extension and relation with literature}

We have proved that the PBS can reproduce at the same time the
bid-ask spread and the volatility smile; but in literature many
authors (see Perignon et al.  \cite{bib:Perignon-Villa}, Renault et al.
\cite{bib:Renault-Touzi} and Rubinstein
\cite{bib:Rubinstein}) have remarked that, generally, the volatility
presents a skewed structure (the graph of implied volatility is
downward sloping); besides in this article we have limited the study
of the PBS model at the martingale case, when the extra returns of
the stock are zero; however, a simple argument of risk aversion
induce leads us to suppose that the parameter $\mu$ (the
extra-returns term in the BS model) must be positive. In a next
paper we will show that the extra returns term has an impact in PBS
model, contrary to BS model, and we can use this effect to generate
a slope in implied volatility. We want emphasize that the PBS model
uses a perturbative approach, i.e. we start with a simple model (the
Black Sholes model) and we adjust the model at the market data
through a perturbation of the principal parameter (the volatility).

Clearly in literature some authors introduce a perturbative approach
in finance, we recall the papers of Hagan et al. \cite{bib:Hagan}
and the book of Fouque et al. \cite{bib:Fouque}; our approach is
however different, it is a probabilistic approach and it can relate
the bid-ask spread and the volatility smile through the simple
economic argument of the existence of uncertainty in the market.

The study of SABR model, a stochastic volatility model introduced by
Hagan et al., is based on a small volatility of
volatility expansion. Now the principal difference between our
perturbation method and the Hagan's one is that our model is based
on a probabilistic point of view, therefore our perturbative
approach takes into account the second order derivative and, in
particular, it can estimate the bias induced by non-linearity. In
the paper of Hagan et al. the approach is based on analysis, so that
they implicitly suppose that an exact value for
 the volatility and, by consequence, for the option price exists,
 their perturbative approach is therefore a first order formulation; they
 cannot then
 justify the presence of a bid-ask spread directly through their model.

 Moreover Hagan et al. start with a more complex model
 (a stochastic volatility model with a volatility driven by an brownian
  diffusion with an independent component) and  they simplify their model with
 the perturbative approach in order to find a closed form,
 therefore in SABR model the perturbative approach is a computation
 tool.
 Our approach allows to justify the perturbative approach with an
 economics reason, the problem for the traders to estimate the value of
 volatility.
 Therefore in our method the perturbation is naturally a consequence of uncertainty on
 volatility, and it has a "physics" nature; finally our model
 start by the classic, well-known and
 accepted Black Scholes model, in one phrase the SABR model is a downward
 perturbative approach the PBS is an upward one.
The principal advantage of the SABR model with respect to PBS model
is that the SABR model can predict the dynamics of the smile when
the forward price of stock changes; this capacity is related with
the parameter $\beta$, and has no relation with the perturbative
approach.

In the book of Fouque, Papanicolau and Sircar the authors study a
model based on the Black-Scholes model with a stochastic volatility
with a fast mean reversion, in this case the perturbative approach
is based on the presence of two time scales the mean reversion time
and the maturity of options, the first is very small compared the
other time scaled, therefore, in their book, Fouque et al. introduce
an expansion of the solution of their prices PDE (based on a
hierarchical PDE system). Therefore, in their book, Fouque et al.
use a perturbative approach with a "physical" nature, i.e. the
presence of a two time scales; but their approach is based on
analysis and it is not (directly) related with an uncertainty on
volatility; finally the Fouque's approach cannot reproduce a bid-ask
spread.

\section{Risk Aversion}\label{app:utility}

In this section we make some recalls on the theory of utility
functions and we show a connection with Error Theory using Dirichlet
Forms. We consider a utility function $U(x): \mathbb{R} \rightarrow
\mathbb{R}$ and $U(x) \in C^2$: let $\rho$ be defined by the
following relation

\begin{equation}\label{avversion-1}
    \mathbb{E}\left[U(X)\right] = U\left(\mathbb{E}\left[X\right] - \rho \right).
\end{equation}

We find, in the case of a small variance, that:

\begin{equation}\label{avversion-absolute}
    \rho = - \frac{\sigma^2}{2}
    \frac{U"\left(\mathbb{E}\left[X\right]\right)}{U'\left(\mathbb{E}\left[X\right]\right)}=
    \frac{\sigma^2}{2} r_a\left(X\right).
\end{equation}

In a similar way we define $\hat{\rho}$ so that:

\begin{eqnarray}
  \mathbb{E}\left[U(X)\right] & = & U\left\{\mathbb{E}\left[X\right]\left(1 - \hat{\rho}
  \right)\right\}: \; \; and \; \; we \; \; find \nonumber \\
  \hat{\rho} &=&  - \frac{\sigma^2}{2}
    \mathbb{E}\left[X\right] \frac{U"\left(\mathbb{E}\left[X\right]\right)}{U'\left(\mathbb{E}\left[X\right]\right)}=
    \frac{\sigma^2}{2} r_r\left(X\right). \label{avversion-relative}
\end{eqnarray}

We can name the three objects:

\begin{enumerate}
    \item{$\rho$ \it is the risk price;}
    \item{ $r_a\left(X\right)$ \it is the absolute index of aversion at the
    wealth X;}
    \item{$r_r\left(X\right)$ \it  is the relative index of aversion at the
    wealth X.}
\end{enumerate}

We write the relations of error and bias, given an error structure on $X$:

\begin{eqnarray*}
  A\left[U(X)\right] &=& U'(X) A\left[X\right] + \frac{1}{2} U"(X) \Gamma\left[X\right] \\
  \Gamma\left[U(X)\right] &=& \left(U'(X)\right)^2 \Gamma\left[X\right]
\end{eqnarray*}

We observe that  the bias of $U(X)$, $A[U(X)]$ is zero if and only
if $A[X] = \rho$; in this case we find

\begin{eqnarray}
  A[X] &=& \frac{r_a}{2} \Gamma[X] \nonumber \\
  \frac{A[X]}{X} &=& \frac{r_r}{2} \frac{\Gamma[X]}{X^2}.
\end{eqnarray}

We have found a relation between the utility function
theory and the error calculus using Dirichlet Forms.
We suppose that all traders buy and sell according to their risk
aversion, and this aversion is represented via a utility function
$U(x)$; then the utility function is defined by its relative index
of aversion.

We make an hypothesis:

\medskip

 Hypothesis $(\ast)$: {\it for a trader the bias of the utility of a
traded wealth
  vanishes.}

\bigskip

We can interpret this hypothesis from an economics point of view in two ways:

\begin{enumerate}

\item the utility function of trader, supposed to be known, is like a lens  traders look
the market through. Traders don't add any effect to
balance their aversion;

\item the vector $(X, \, \mathcal{A}, \, \Gamma)$ is supposed to be known, we can define the utility function of a trader as
function $U(x)$ that cancels the bias of $U(X)$ where X is the
considered wealth.
\end{enumerate}

Under hypothesis $(\ast)$ we have two relations between the utility
function and Dirichlet Forms:

\begin{enumerate}
\item{$A[X] = \rho(X)$, where $\rho$ is the risk price;}
\item{ $r_r(X) = 2\frac{X A[X]}{\Gamma[X]}$}
\end{enumerate}

\begin{remarque}
  Thanks to relation \ref{eqn:particular-case} we can define the class of utility function
  that preserve vanishing bias and slope at the money.

  \begin{eqnarray*}
    r_r\left(\sigma \sqrt{T} \right) & = & \frac{\sigma^2 T}{4} \\
    X\frac{U''(X)}{U'(X)} & = & - \frac{X^2}{4} \\
    U'(X) & = & e^{-\frac{X^2}{8}} \\
    U(X) & = & \mathcal{N}\left(\frac{X}{2}\right)
  \end{eqnarray*}

  Where $\mathcal{N}(X)$ is the distribution function of the normal law.
This utility function is concave if the wealth is positive and
convex otherwise.

  \end{remarque}

\section{Conclusion and Economics Interpretation}

In this paper we have studied the impact of a perturbation on
volatility in Black-Scholes model in absence of drift and term
structure; in particular we have dealt with the problem of call hedging.

We have proposed a new model for option pricing, called Perturbed
Black-Scholes model; the basic idea is take into account the effect
of uncertainty of volatility value in order to reproduce, at the
same time, the spread bid-ask and the smile on implied volatility;
the mainly advantage of this model is that it is based on the
classic Black Scholes model and the price of an option in PBS model
is the BS price plus a small perturbation that depends only on the
greeks founded with BS formula, therefore the computation of PBS
price is given by a closed form.

The PBS model depends on four parameters, naturally on the
volatility of stock; but also on the variance of the estimated
volatility $\, \epsilon
 \, \Gamma[\sigma](\sigma_0)$,  on a relative index $r_r(\sigma)$, that
represents the ratio between the  bias and the variance  of the
estimated volatility, and, finally, contrary to Black Scholes model,
on the drift $\mu$ that represents the extra returns of stock; the
impact of the drift rate $\mu$ will be studies in a next paper.

In particular, if we set the relative index to be equal to
$\frac{1}{4}\sigma_0^2 \,T$, we have proved that the implied
volatility present a smile around the money.

Finally we have defined a Local Volatility Model equivalent
to Perturbed Black
 Scholes Model as far as vanilla options are concerned; the related
  local volatility function is defined by Dupire formula:

\begin{equation*}
  \sigma^2(T, \, K) \approx \sigma_0^2 \left\{ 1 + \frac{\epsilon}{2}\left[ 4 \frac{A\left[\sigma \sqrt{T}\right]}{\sigma_0 \sqrt{T}} - \left[\sigma^2 T +
       2 - \frac{4 \ln\left(\frac{x}{K}\right)^2}{\sigma_0^2 T} \right] \frac{\Gamma\left[\sigma
         \sqrt{T}\right]}{\sigma_0^2 T} \right] \right\}
\end{equation*}

\end{document}